\documentclass[prl,twocolumn,superscriptaddress,aps]{revtex4-1}
\usepackage{ulem}
\usepackage{bbm}
\usepackage{color}
\usepackage{graphicx}
\usepackage{physics}
\usepackage{hyperref}
\usepackage{comment}
\usepackage{MnSymbol,wasysym}
\usepackage{siunitx} 
\usepackage{appendix}
\usepackage{verbatim}
\usepackage{subfigure}
\usepackage{fullpage}

\newcommand{\commentout}[1]{}
\newcommand{\OP}{\Omega_p (t)}
\newcommand{\OS}{\Omega_S(t)}

\begin{document}

\title{Tailoring population transfer between two hyperfine ground states of $^{87}$Rb}

\author{Aleksandra Sierant}
\email{aleksandra.sierant@icfo.eu}
\affiliation{\JUAddress}
\affiliation{\ICFOAddress}

\author{Marek Kopciuch}
\affiliation{\JUAddress}

\author{Szymon Pustelny}
\affiliation{\JUAddress}

\date{August 2022}

\newcommand{\JUAddress}{Institute of Physics, Jagiellonian University in
Krak\'ow, \L{}ojasiewicza 11, 30-348 Krak\'ow, Poland}

\newcommand{\ICFOAddress}{ICFO - Institut de Ciencies Fotoniques, The Barcelona Institute of Science and Technology, Castelldefels (Barcelona) 08860, Spain}

\begin{abstract}
In this paper, we investigate the coherent control over a complex multi-level atomic system using the stimulated Raman adiabatic passage (STIRAP). Based on the example of rubidium-87 atoms, excited with circularly-polarized light at the D$_1$ line, we demonstrate the ability to decompose the system into three- and four-level subsystems independently interacting with light beams. Focusing on the four-level system, we demonstrate that the presence of an additional excited state significantly affects the dynamics of the system evolution. Specifically, it is shown that, through the appropriate tuning of the light beams, some of the transfer channels can be blocked, which leads to better control over the system. We also demonstrate that this effect is most significant in media free from inhomogenous broadening (e.g., Doppler effect) and deteriorates if such broadening is present. For instance, motion of atoms affects both the efficiency and selectivity of the transfer.
\end{abstract}

\maketitle

\section{I. Introduction} \label{intro}

The preparation of a system in a desired quantum state plays a crucial role in the advancement of quantum technologies \cite{acin2018quantum}, ranging from quantum computation and simulations \cite{PhysRevLett.87.037901, smith2013quantum, divincenzo2000physical, bloch2012quantum,o2007optical} to quantum cryptography and information processing \cite{gisin2002quantum}. An important step in the development of quantum-information applications is the ability to transfer a quantum state between two different physical manifolds, which, among others, may enable the implementation of quantum memory schemes \cite{lvovsky2009optical,heshami2016quantum}. A particular system in which such a transfer can be realized consists of two long-lived hyperfine ground levels of alkali atoms \cite{phillips2001storage,lukin2003colloquium,hammerer2010quantum,hosseini2011high}.

Stimulated raman adiabatic passage (STIRAP) \cite{gaubatz1990population,bergmann1998coherent,vitanov2001laser,vitanov2001coherent,bergmann2019roadmap,vitanov2017stimulated,shore2017picturing,bergmann2015perspective} is a powerful technique that allows the adiabatic transfer of the population between two long-lived states.  Originally designed for a three-level system, STIRAP coherently transfers the population between two long-lived lower-energy states $\ket{\mathrm{i}}$ and $\ket{\mathrm{f}}$ by coupling them to a fast-decaying excited state $\ket{\mathrm{e}}$ with a counter-intuitive sequence of the so-called Stokes and pump pulses, respectively coupling the states $\ket{\mathrm{f}}$ and $\ket{\mathrm{e}}$ and the states $\ket{\mathrm{i}}$ and $\ket{\mathrm{e}}$. During the pulse sequence, the population is trapped in a dark state that is a superposition of the $\ket{\mathrm{i}}$ and $\ket{\mathrm{f}}$ states and has a vanishing overlap with the excited state $\ket{\mathrm{e}}$ \cite{bergmann1998coherent}. At the same time, if the Stokes pulse precedes the pump pulse, the dark state initially has a large overlap with the state $\ket{\mathrm{i}}$, but later with the state $\ket{\mathrm{f}}$ (the population of the state $\ket{\mathrm{e}}$ is negligible at any stage of evolution). In turn, the scheme enables a coherent transfer of the population from the state $\ket{\mathrm{i}}$ to the state $\ket{\mathrm{f}}$. Until now, STIRAP has been investigated in many systems, including cold atoms \cite{du2016experimental,gearba2007measurement}, molecules \cite{winkler2007coherent,danzl2008quantum,park2015ultracold}, and ions \cite{sorensen2006efficient,higgins2017coherent}. 

It should be noted that the presence of other excited states may affect the coherent transfer of the population \cite{coulston1992population,vitanov1999adiabatic,fernandez2018adiabatic,yang2012enhancement}. For example, in room-temperature alkali-metal vapors \cite{oberst2007time,kuhn1998coherent,wang2008coherence}, for which the Doppler broadening is comparable to or larger than the excited state splitting, the three-level model is not accurate \cite{tiecke2010properties} and calls for a more elaborate description of the phenomenon \cite{coulston1992population,vitanov1999adiabatic,fernandez2018adiabatic,yang2012enhancement,gong2004complete}.

In this work, we investigate STIRAP for coherent population transfer between two hyperfine ground-state levels of $^{87}$Rb atoms. We show that, despite the rich energy-level structure of rubidium, the system can be
effectively decomposed into three- and four-level subsystems.  We also demonstrate that appropriately polarized and tuned light provides control over the selectivity of the transfer. The control investigated in this paper is qualitatively different from the previous experiments, which required an additional strong perturbation (e.g., large magnetic field \cite{shore1995coherent,martin1995coherent,martin1996coherent}, which induces a strong splitting of the Zeeman sublevels) to achieve the control. However, it should be noted that, despite the specific context, the discussion presented in this paper is generic and can be used for any four-level system. 

The paper is organized as follows. Section II is dedicated to a theoretical description of STIRAP in a specific context of $^{87}$Rb atoms excited at the D$_1$ line. In particular, the so-called local adiabatic condition is introduced in the four-level system. Section III presents results of numerical simulations of population transfer between two long-lived ground states of $^{87}$Rb. This section analyzes the STIRAP efficiency in Doppler-free and Doppler-broadened media, comparing similarities and differences between the two cases. Conclusions are summarized in Sec. IV. Finally, the adiabatic conditions for STIRAP in the four-level system are derived in Appendix. 

\section{II. Theory of STIRAP in a 4-level system} \label{theory}

\begin{figure}[h!]
    \includegraphics[width=\columnwidth]{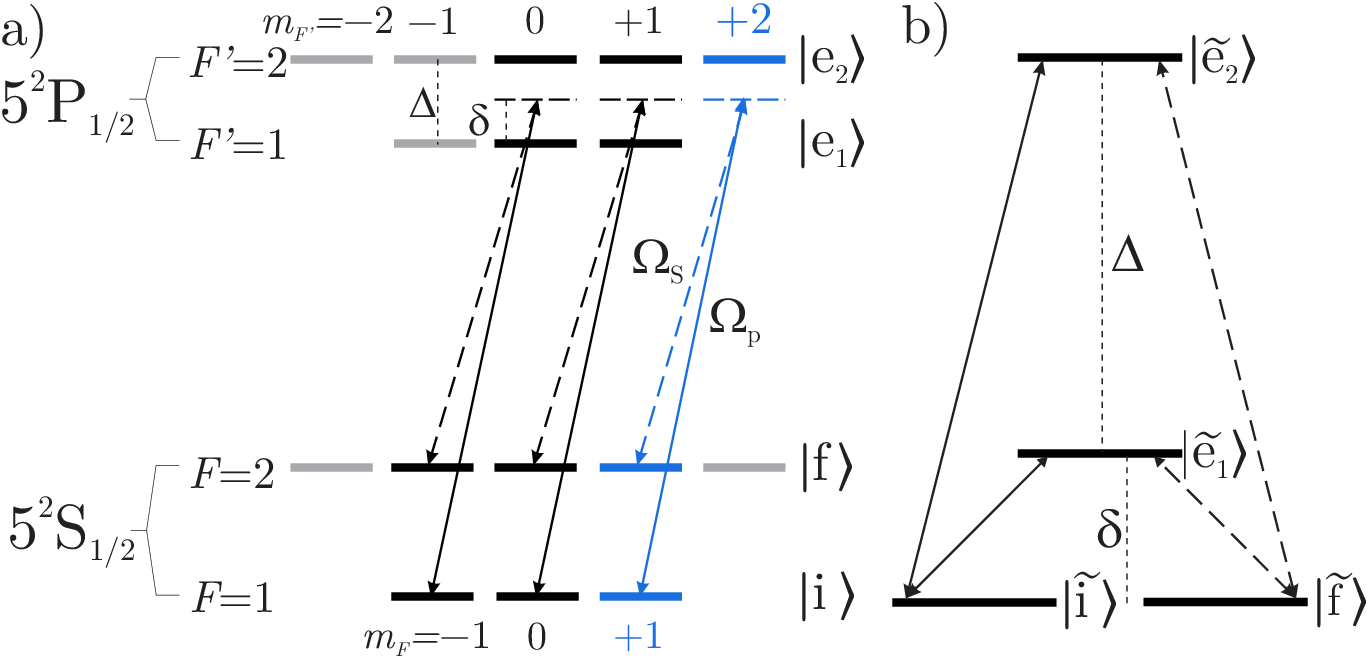}
    \caption{a) Energy-level diagram corresponding to the $^{87}$Rb D$_1$ line along with the transitions induced by the $\sigma^{+}$--polarized pump (solid lines) and Stokes (dashed lines) beams. b) Example of a separated four-level system in the rotated frame (see discussion in the main text).}
    \label{figure1}
\end{figure}

Consider the energy-level structure associated with the $5^2S_{1/2}\rightarrow 5^2P_{1/2}$ transition (the D$_1$ line) in~$^{87}$Rb atoms (Fig.~\ref{figure1}a).  The system has four hyperfine levels (two ground levels and two excited levels) of rich magnetic-sublevel structures.   
To provide the ability to excite only a given pair of states, the spectral widths of the Stokes and pump pulses must be narrower than the splitting of long-lived hyperfine ground states $\ket{\mathrm{i}}$ and $\ket{\mathrm{f}}$, i.e.,
\begin{equation}
     T_{S,p} \gg \dfrac{\hbar}{\Delta E},
     \label{eq:separation}
\end{equation}
where $\Delta E$ is the energy splitting of the ground levels and $T_{S}$ $(T_p)$ is the duration of the Stokes (pump) pulse. Under such conditions, the Stokes pulse does not excite the atoms that reside in the initial state $\ket{\mathrm{i}}$ and the pump light does not excite the atoms present in the final state $\ket{\mathrm{f}}$. To fulfill this condition in the considered case of $^{87}$Rb atoms ($\Delta E/h \approx \SI{6.8}{\giga\hertz} $), the pulses must last at least a hundred picoseconds.

Assuming that the polarizations of the Stokes and the pump beams are the same (here, we considered $\sigma^+$--polarized light), the selection rules indicate that the considered sixteen-level system can be decomposed into three independent subsystems, two of which are four-level systems (in Fig.~\ref{figure1}a marked in black) and one is a three-level system (marked in blue). The remaining five gray-colored states denote the sublevels that do not participate in STIRAP and hence are neglected in our considerations.

Let us now focus on a single four-level subsystem. Using the standard dipole approximation, one can write an explicit form of the Hamiltonian of the system, which, within the rotating-wave approximation (see Fig.~\ref{figure1}b), takes the form
\begin{equation}
\small
     H = \hbar\left(\begin{array}{cccc}
      \delta & 0 & C_{i e_{1}} \Omega_{p}(t) & C_{i e_{2}} \Omega_{p}(t)\\
      0 & \delta & C_{f e_{1}} \Omega_{S}(t) & C_{f e_{2}} \Omega_{S}(t)\\
      C_{i e_{1}}^{*} \Omega_{p}(t) & C_{f e_{1}}^{*}\Omega_{S}(t) & 0 &
0 \\
      C_{i e_{2}}^{*} \Omega_{p}(t) & C_{f e_{2}}^{*}\Omega_{S}(t) & 0 &
\Delta
     \end{array}\right).
     \label{eq:hamiltonian}
\end{equation}
Here $\delta$ is the one-photon detuning, $\Delta$ is the frequency splitting of the excited levels, and $C_{x y}$ denotes the coupling constant between the states $\ket{\tilde{\mathrm{x}}}$ and $\ket{\tilde{\mathrm{y}}}$ given in the rotated basis (see Appendix). In the considerations, light beams satisfy the two-photon resonance condition ($\omega_S-\omega_p=\Delta E/\hbar$, where $\omega_S$ and $\omega_p$ are the carrier frequencies of the Stokes and pump beams, respectively), and $\Omega_{S}(t)$ [$\Omega_{p}(t)$] is the slowly-varying Rabi frequency of the Stokes (pump) beam (for more details see Appendix).

In general, diagonalization of a four-level system leads to eigenstates that are non-trivial superpositions of all four unperturbed states. However, there are two cases for which the STIRAP-generated dark state is similar to the dark state of conventional three-level STIRAP \cite{vitanov1999adiabatic}. The first case (I) occurs when the couplings of both excited states to the corresponding ground states are equal up to the sign, i.e., $C_{ie_{1}}=\pm C_{ie_{2}}$ and $C_{fe_{1}}=\pm C_{fe_{2}}$. The second case (II) is associated with the situation where the ground states are coupled equally to the corresponding excited states, i.e., $C_{ie_{1}}=\pm C_{fe_{1}}$ and $C_{ie_{2}}=\pm C_{fe_{2}}$. Due to the Clebsch-Gordan coefficients, these conditions can be achieved in $^{87}$Rb only for the same circular polarization of both beams. If these conditions are met, the four-level dark state is given by the standard formula \cite{bergmann1998coherent}
\begin{equation}
    \ket{\delta} = \cos \vartheta \left(t\right) \ket{\tilde{\mathrm{i}}} - \sin \vartheta \left(t\right) \ket{\tilde{\mathrm{f}}},
    \label{eq:DarkStateGeneric}
\end{equation}
where $\vartheta$ is the mixing angle determined by
\begin{equation}
   \tan \vartheta\left( t \right) = \dfrac{\Omega_{p}(t)}{a\Omega_{S}(t)},
   \label{eq:mixing_angle}
\end{equation}
and $a=\left( C_{ie_{1}}/C_{fe_{1}} \right)^{*}$.
\begin{figure}
    \includegraphics[width=1\columnwidth]{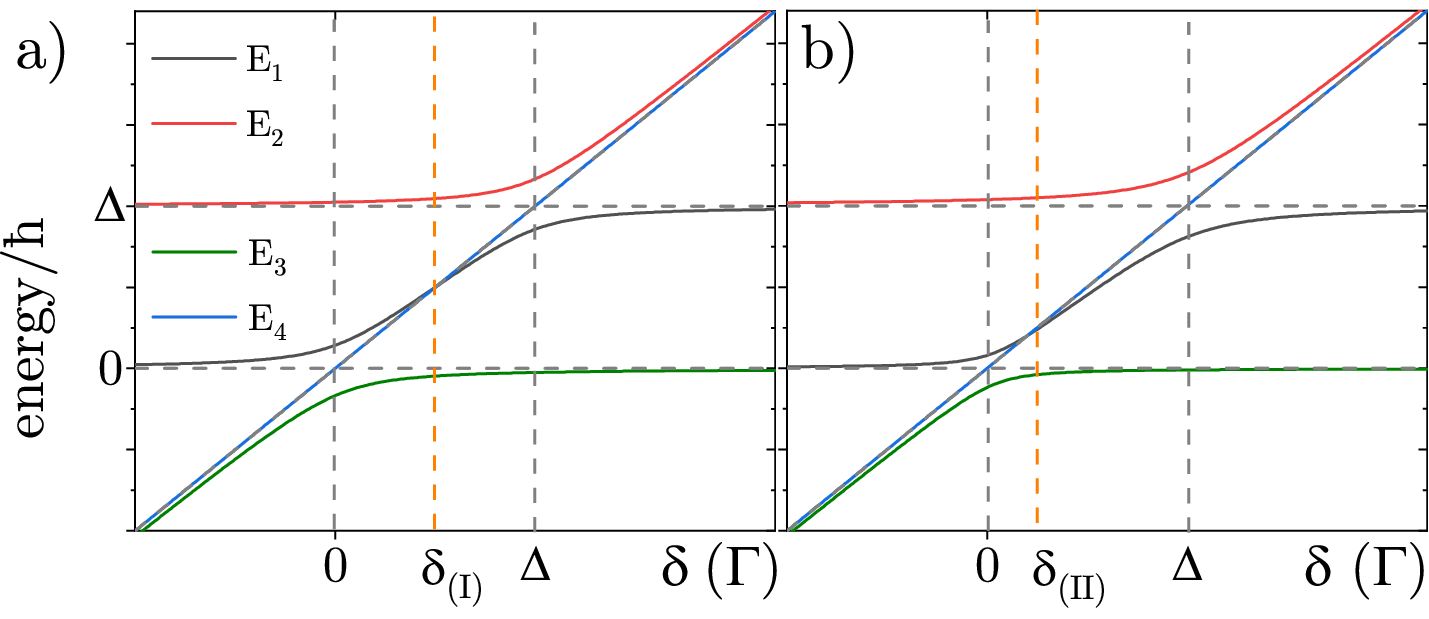}
    \caption{Eigenenergies of system as functions of the one-photon detuning $\delta$ for equal pump and Stokes Rabi frequencies, $\Omega_S(t)=\Omega_p(t)$. Couplings are equal for both ground  (a)  and  both excited (b) states.  Solid lines denote the eigenvalues of the Hamiltonian \eqref{eq:hamiltonian}. The blue line corresponds to the dark state, whereas the dashed gray lines correspond to the bare energies of states $\ket{\tilde{\mathrm{i}}},\ket{\tilde{\mathrm{f}}}, \ket{\tilde{\mathrm{e}}_1},\ket{\tilde{\mathrm{e}}_2}$. Dashed orange line indicates the intersection of the curves.}
    \label{fig:Eigenvalues}
\end{figure}
The corresponding eigenenergy of the dark state is equal to $\hbar \delta$, where $\delta$ is one-photon detuning. In the rotated basis, the energies of the excited states $\ket{\tilde{\mathrm{e}}_{1}}$ and $\ket{\tilde{\mathrm{e}}_{2}}$ are $0$ and $\hbar\Delta$, respectively. The coupling between the ground and excited states modifies the eigenstates and significantly shifts the eigenenergies. The coupling of each excited state with both ground states leads to avoided crossings at $\delta=0$ and $\delta=\Delta$.  The linear dependence of the dark state on $\delta$ results in a level crossing of the dark state with one of the eigenstates, provided by the lack of coupling between the ground levels. The crossing occurs between 0 and $\Delta$, $0<\delta<\Delta$, see Fig.~\ref{fig:Eigenvalues}. In the two cases (I) and (II) described above, the crossing positions take different values. Specifically, when both excited states are equally strong coupled to the ground states [case (I)] both avoided crossings have the same widths and the level crossing occurs exactly in the middle between the two avoided crossings $\delta_{(\textrm{I})}=\Delta/2$ (Fig.~\ref{fig:Eigenvalues}a).  However, in the other case, the avoided crossings have unequal widths due to unequal coupling strengths, and the position of the level crossing is given by $\delta_{(\textrm{II})} = \Delta\vert C_{ie_1}\vert^2/( \vert C_{ie_{1}}\vert^{2}+\vert C_{ie_{2}}\vert ^{2})$. These two cases are shown in Fig.~\ref{fig:Eigenvalues}.

The adiabaticity of the standard three-level STIRAP is guaranteed by global and Local Adiabatic Conditions (LAC) \cite{kuklinski1989adiabatic, vitanov2017stimulated}. The presence of the fourth state modifies the eigenstates of the system, so that the standard adiabatic conditions are no longer valid. In Appendix, we derived modification of LAC, taking into account the existence of the additional excited state. The condition takes the form
\begin{equation}
\small
     \text{min}_{\mathrm{y}\neq \delta} \left\vert E_{\mathrm{y}} - \hbar\delta \right\vert \gg
\hbar \vert a \vert \dfrac{\vert  \Omega_{p}(t) \dot\Omega_{S} (t) -
\Omega_{S}(t) \dot\Omega_{p} (t)\vert}{\Omega_{p}^{2} (t)+ \vert a
\vert^{2} \Omega_{S}^{2} (t)},
     \label{eq:localCondition}
\end{equation}
where $\mathrm{y}$ indicates all eigenstates except the dark state. The condition states that the Stokes and pump pulse smoothness and their overlap are upper-bounded by the minimal separation between the dark state and closest-laying eigenstate. Since in the considered case there is a level crossing, the left-hand side of Eq.~$\eqref{eq:localCondition}$ is equal to 0. This implies that regardless of pulse smoothness, LAC is not fulfilled, leading to a significant deterioration of the population-transfer efficiency at $\delta_{(\text{I})}$ or $\delta_{(\text{II})}$. 

It should be finally stressed that although identical results can be achieved for $\sigma^{-}$--polarized light (polarization $\pi$ is outside our interest due to the existence of a forbidden transition between the magnetic sublevels of $m_F=0$ of the states of the same total angular momentum, $F=F'$). The approach does not work for different pump- and Stokes-beam polarizations. In this case, there is no eigenstate that has a vanishing overlap with any of the excited states. Thereby, the excited states are always partially populated, which, through spontaneous emission, compromises coherence of the transfer. Yet, it can be shown that even in this case there is an eigenstate which, asymptotically in time, behaves like the STIRAP dark state \cite{vitanov1999adiabatic}. That is, for $t\rightarrow -\infty$ behaves like $\ket{i}$, and for $t\rightarrow \infty$ transforms to $\ket{f}$.

\begin{figure*}[htb!]
\includegraphics[width=2\columnwidth]{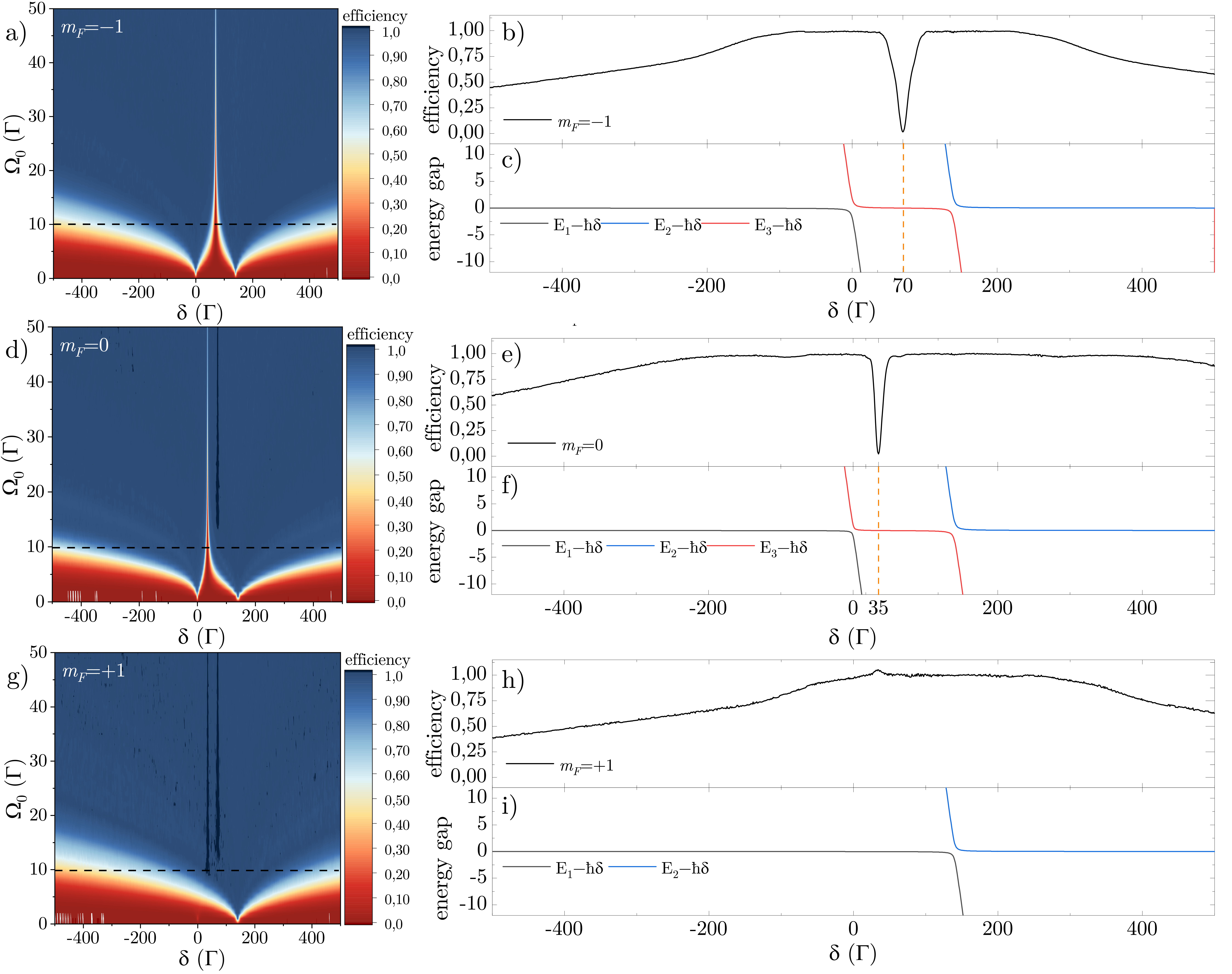}
\caption{Efficiency of population transfer from the m$_F$=-1 (a,b,c), m$_F$=0 (d,e,f) and m$_F$=+1 (g,h,i) sublevels of the $F=1$ state of the full manifold of states of $^{87}$Rb D$_1$ line. (a,d,g) Efficiency of the population transfer versus Rabi-frequency amplitude $\Omega_0$ and one-photon detuning $\delta$ in the natural-linewidth units. (b,e,h) Cross-sections through the maps at Rabi-frequency amplitude 10$\Gamma$, marked by a line. (c,f,i) Energy separation E$_i-~\hbar\delta$ as a function of the one-photon detuning $\delta$. Vertical dashed lines indicate the positions of the minimum transfer efficiency, and the energy gap is given in arbitrary units.}
\label{figure3}
\end{figure*}

\section{III. Results and Discussion} \label{results}

\begin{figure}
\includegraphics[width=\columnwidth]{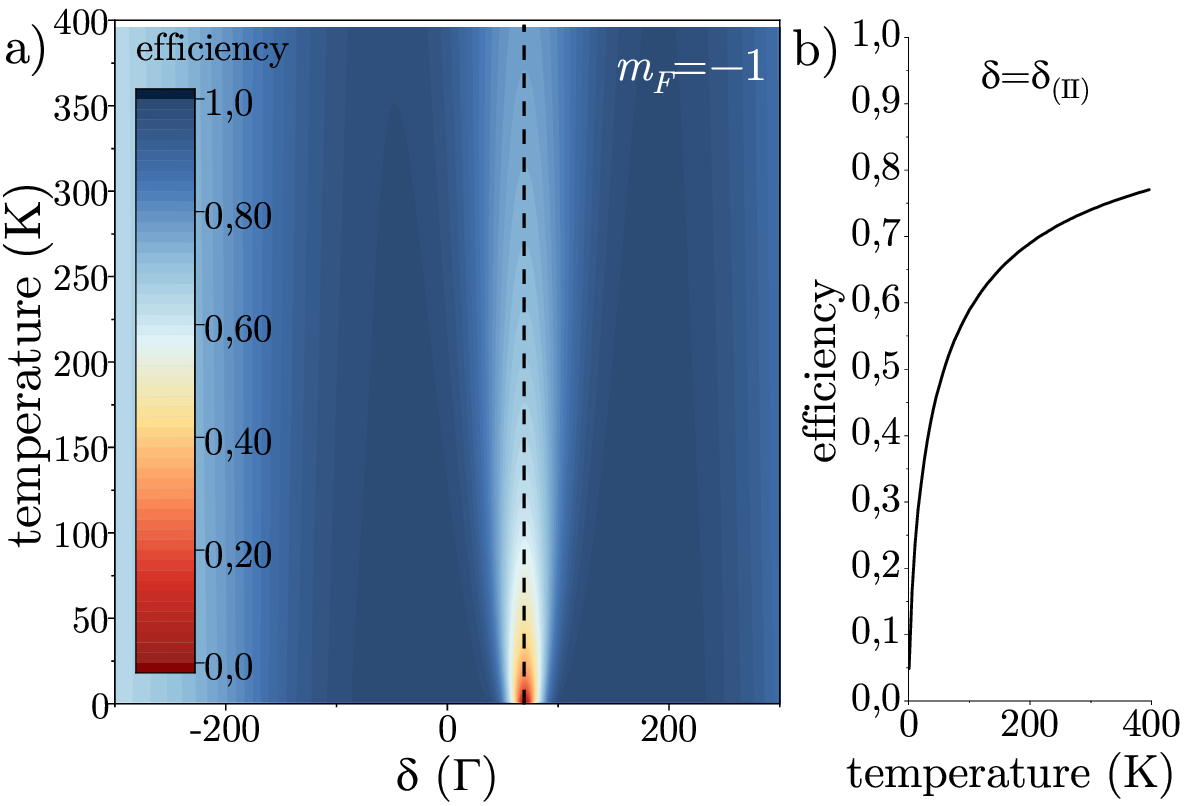}
\caption{a) Population transfer calculated for subsystem of m$_F$=-1 as
a function of the one-photon detuning $\delta$ and temperature at peak
Rabi frequency 10$\Gamma$. b) Population of the final state at
$\delta=\delta_{(II)}$ versus temperature (the position is denoted by
black, dashed line).}
\label{figure4}
\end{figure}

In this section, we investigate the transfer of population between two hyperfine ground states of $^{87}$Rb. While in the rotating frame the states are degenerate, they can still be individually addressed by the Stokes and pump beams. The excited states $\ket{\mathrm{e}_1}$ ($F'=1$) and $\ket{\mathrm{e}_2}$  ($F'=2$) are separated by $\Delta=140\Gamma$, where $\Gamma$ is the relaxation rate of the excited state. Each of the states is characterized by an additional magnetic-sublevel structure. For the simulations, we assume equal population of all sublevels of the $F=1$ state (the state $\ket{\mathrm{i}}$) and no population in the $F=2$ state (the state $\ket{\mathrm{f}}$), nor in any of the excited states. The system interacts with two Gaussian-shaped light pulses, $\Omega_{S,p}(t)=\Omega^0 \exp[(t-\tau_{S,p})^2/(2T^2)]$, where $\tau_{S,p}$ are the delays of the Stokes and pump pulses, respectively, $\Omega^0$ is the amplitude of the pulses, and $T$ is their duration. The arbitrarily chosen pulse duration is $T=300\frac{1}{\Gamma}$ for both pulses ($T=T_S=T_p$) and the separation of the pulses is $200\frac{1}{\Gamma}$. In our analysis, we consider the same circular polarization of the Stokes and pump beams ($\sigma^+$ polarization). Evolution of such a sixteen-level system is calculated by numerical solution of the master equation \cite{tannoudji1992atom,auzinsh2010optically}.

Figure~\ref{figure3} presents the efficiency of population transfer in three subsystems specified in the previous section, i.e., two four-state subsystems (first involving the $m_F=-1$ ground-state sublevels and second the $m_F=0$ sublevels), and the three-level subsystem, accounting for the $m_F=1$ sublevels. Transfer efficiency is defined as the ratio of the final population of the state $\ket{\tilde{\mathrm{f}}}$ to the initial population of the state $\ket{\tilde{\mathrm{i}}}$.  The color maps show the efficiency as a function of one-photon detuning $\delta$ and the Rabi-frequency amplitude $\Omega^0$. Regardless of the subsystem, failure of STIRAP can be observed at low Rabi-pulse amplitudes. This stems from the violation of the global adiabatic condition (light power is too weak to generate a strong superposition of ground and excited levels) \cite{vitanov2017stimulated}.  For a higher Rabi frequency, only a single maximum is observed in the three-level system in the transfer efficiency measured versus the one-photon detuning. At the same time, in the four-level subsystems, where an additional excited level is present, two maxima are observed at the $0$ and $140\Gamma$ detunings. Each of them corresponds to STIRAP involving different excited states. Interestingly, however, the transfer efficiency drops almost to zero for a specific detuning between the two maxima. This is clearly visible in the cross sections shown in Figs.~\ref{figure3}b and e. The positions of the minima at $\delta=70\Gamma$ (Fig.~\ref{figure3}b) and $\delta=35\Gamma$ (Fig.~\ref{figure3}e) are determined by violation of LAC and agree well with the discussion presented in Sec.~II. Moreover, for the case presented in Fig.~\ref{figure3}e, one can see that the efficiency plateau is not symmetric with respect to the $0<\delta < \Delta$ region. Specifically, the plateau extends further toward positive detunings due to the stronger avoided crossing at $\delta=\Delta$ than at $\delta=0$ (case II). Another interesting feature visible in the maps are the dark blue vertical stripes observed in shown in Figs.~\ref{figure3}d and g. The stripes indicate the transfer efficiency higher than unity. Although this excess may appear at first sight to be wrong, it is a result of additional repopulation of the level via spontaneous emission from different excited states (incoherent pumping). In fact, this process is present when LAC is violated, and a non-negligible population arises in the excited states. At the same time, the sublevel of $m_F=-1$ is repopulated only by itself \cite{tannoudji1992atom}, so there is no excess of the population relative to the initial-state population.

For a better understanding of the detuning dependences, Figs.~\ref{figure3}c, f, and i show the energy difference between the dark state and other eigenstates. As shown, one of the eigenstates (solid red line) crosses with the dark state (dashed orange line) at $70\Gamma$ (Fig.~\ref{figure3}c) and $35\Gamma$ (Fig.~\ref{figure3}f). This clearly demonstrates the relationship between the reduction in transfer efficiency and the violation of LAC [Eq.~\eqref{eq:localCondition}].  

Reduction of the efficiency of population transfer in a specific subsystem for given detunings can be used for control over the population transfer, even when resonant pulses are used. Specifically, by tuning the light, one can effectively ``turn off'' the transfer from the specific.  For example, using the $\sigma^{+}$--polarized light of $\delta=\delta_{(\text{I})}$ ($\delta=\delta_{(\text{II})}$) enables one to block the transfer from the $m_F=-1$ ($m_F=0$) magnetic sublevel.  In the same manner, $\sigma^{-}$--polarized light allows to 'block' the transfer from the $m_F=1$ sublevel. So far, this kind of control and selectivity in the transfer has required the use of a magnetic field, which splits the Zeeman sublevels so strongly that they can be independently addressed \cite{shore1995coherent,martin1995coherent,martin1996coherent}. With our technique, the transfer is controlled even if the magnetic sublevels are degenerate.

The dependence of population-transfer efficiency on the one-photon detuning $\delta$ plays an important role in room-temperature atomic vapors. In such systems, the Doppler broadening of the transition is about two orders of magnitude larger than the excited-state relaxation rate (natural width). In turn, the overall efficiency of STIRAP is affected by atoms that are Doppler-tuned to $\delta_{(\text{I})}$ or $\delta_{(\text{II})}$. Figure~\ref{figure4} presents the transfer efficiency to the $m_F=-1$ sublevel versus one-photon detuning $\delta$ and rubidium-vapor temperature. As shown above for $\delta=\delta_{(\text{II})}$, the efficiency suffers from the LAC violation at low temperature and no transfer is observed. However, the efficiency of the transfer increases with temperature (see Fig.~\ref{figure4}b), which limits the control over the system. On the other hand, the maximum transfer efficiency (efficiencies integrated over the velocity distribution) decreases with temperature. For instance, the efficiency for the room-temperature vapor, interacting with light of Rabi amplitude $10\Gamma$ and zero detuning ($\delta=0$), decreases from 99.9$\%$ to 93.5$\%$ compared to the cold atomic ensemble. This behavior may be a problem in the applications of STIRAP in quantum-state engineering with hot atomic vapors.

\section{IV. Conclusions} \label{conclusions}

In summary, we theoretically investigated coherent population transfer between two long-lived ground states of $^{87}$Rb excited at the D$_1$ line. The transfer is based on the well-established STIRAP method. With our analysis, we showed that the sixteen-level system of rubidium can be decomposed into three- and four-level independent subsystems. Among them, of particular interest, was the four-level system. The efficiency of the transfer between two ground states was investigated with respect to parameters such as Stokes/probe-pulse amplitude and one-photon detuning. With our analysis, we demonstrated control over the transfer of population from specific magnetic sublevels. Such a control is achieved by tuning the light to the specific regions where the local adiabatic condition is not fulfilled. 

In contrast to previous approaches, our technique does not require the use of additional magnetic fields or a strong light beam. This opens interesting possibilities for tailoring complex quantum states. In this context, the ability of blocking the population transfer from a particular Zeeman sublevel can be beneficial for the generation of complex states involving two hyperfine levels. It should also be noted that the selectivity is partially lost in room-temperature atomic vapors, where Doppler broadening prevents adiabatic condition violation. 

Finally, the presented model can be used to consider the transfer of coherences between two ground states. This may be of particular interest for systems with long-lived coherence \cite{pustelny2011tailoring, Davidson1995Long}, where the technique may allow the extension of a manifold for quantum-state manipulation. This approach and its applicability for quantum-state manipulation could be verified using quantum-state tomography \cite{Deutsch2010Quantum, Kopciuch2022Optical}.

\section{Acknowledgements}
The authors thank Nikolai Vitanov for the stimulating discussions. This project was supported by the National Science Center, Poland, under the grant 2019/34/E/ST2/00440 and 2020/N17/MNW/000015. A.S. acknowledges support from the Government of Spain (Severo Ochoa CEX2019-000910-S), Fundació Cellex, Fundació Mir-Puig, and Generalitat de Catalunya (CERCA, AGAUR). M.K. acknowleges the "Research support module" as part of the "Exellence Initiative -- Research University".

\section{Appendix} \label{appendix}
\renewcommand{\theequation}{A.\arabic{equation}}
\setcounter{equation}{0}

\subsubsection{Hamiltonian of the system}

We consider a four-level system with two hyperfine ground levels and two hyperfine excited levels. The electric fields of the Stokes and pump beams are given by
\begin{equation}
    \mathbf{E_{S,p}}(t) = \boldsymbol{\mathrm{\epsilon}} E^0_{S,p}f_{S,p}(t) \cos \left( \omega_{S,p} t \right),
\end{equation}
where $E^0_{S}$ $(E^0_{p})$ is the electric-field amplitude of the Stokes (pump) light pulse,  $f_{S}(t)$ $[f_{p}(t)]$ is the slowly-varying Gaussian envelope of the pulses with normalized amplitudes, $\omega_{S}$ $(\omega_{p})$ is the Stokes (pump) light frequency, and $\boldsymbol{\mathrm{\epsilon}}$ is the light polarization vector that is identical for both beams. We use the standard dipole approximation, with the interaction term \mbox{$V_E=-\mathbf{E}\cdot\mathbf{d}$}, where $\mathbf{d}$ is the electric dipole moment. The Hamiltonian can be converted to the rotating frame by using a rotation generator given by $\text{diag} \left(\omega_{p}, \omega_{S}, 0, 0 \right) $. We assume that both ground and both excited states are coupled with the beams and two-photon resonance is fulfilled, i.e., 
\begin{equation}
    \omega_S-\omega_p = \dfrac{\Delta E}{\hbar},
\end{equation}
where $\Delta E$ is the energy difference between the initial and final states.  In this case, the off-diagonal elements of the Hamiltonian contain both stationary and oscillatory terms. In rotating-wave approximation (RWA) we neglect the terms oscillating at $2\omega_{p}$, $2\omega_{S}$ and $\omega_{p}+\omega_{S}$, which allows us to write the Hamiltonian as
\begin{widetext}
\begin{equation}
    H=
     \hbar \mathcal{C} \bullet \left\lbrace
     \left(\begin{array}{cccc}
         \delta & 0 & \OP & \OP\\
         0 & \delta & \OS & \OS\\
         \OP & \OS & 0 & 0 \\
         \OP & \OS & 0 & \Delta 
    \end{array}\right) 
    +\left[ 
        e^{-i \left(\Delta E / \hbar\right)  t}\left(\begin{array}{cccc}
        0&0&0&0\\
        0&0&\OP&\OP\\
        \OS&0&0&0\\
        \OS&0&0&0
    \end{array} \right) + h.c.
    \right] \right\rbrace,
    \label{eq:timeDependantHamiltonian}
\end{equation}
\end{widetext}
where $\hbar \Delta$ is the energy difference between the excited states, $\delta$ is the one-photon detuning, and $\Omega_{S}(t)$ $[\Omega_{p}(t)]$ is the slowly-varying Rabi frequency of Stokes (pump) beam given by $ \Omega_{S,p} (t) = E^0_{S,p} f_{S,p}(t)/ (2\hbar) \langle 1/2 \Vert \mathbf{\hat{d}} \Vert 1/2 \rangle$, with the reduced matrix element $\langle 1/2 \Vert \mathbf{\hat{d}} \Vert 1/2 \rangle$ defined as in Ref.~\cite{sobelman2012atomic}. Note that the reduced matrix element is expressed by the quantum number $J$, which is convenient when the excitation of alkali-metal atoms is considered on the D$_1$ line. $\mathcal{C}$ is the coupling-constant matrix
\begin{equation}
\small
    \mathcal{C} = \left(\begin{array}{cccc}
     0 & 0 & C_{1m_F,1m_{F'}} & C_{1m_F,2m_{F'}} \\
     0 & 0 & C_{2m_F,1m_{F'}} & C_{2m_F,2m_{F'}} \\
     C_{1m_F ,1m_{F'}}^{*} & C_{2m_F ,1m_{F'} }^{*} & 0 & 0 \\
     C_{1m_F ,2m_{F'}}^{*} & C_{2m_F ,2m_{F'} }^{*} & 0 & 0
    \end{array}\right)
    \label{eq:SIhamiltonian}
\end{equation}
\normalsize
with $\bullet$ representing the element-wise / Hadamard product ($\left[ A \bullet B\right]_{ij} = A_{ij}B_{ij}$). The coupling-constant matrix elements $C_{Fm_{F}F'm_{F'}}$ are given by
\begin{align}
    C_{Fm_F ,F'm_{F'}}&=(-1)^{3/2+I+F'} \sqrt{(2F+1)} \times \nonumber \\
    & \times \braket{F'm_{F'}}{1qFm_F} \left\{ \begin{array}{ccc}
    1/2 & F & I \\
    F' & 1/2 & 1 \\
    \end{array} \right\},
\end{align}
where $q$ is the index of light polarization in the spherical basis and $I$ is the nuclear quantum number. Since the RWA Hamiltonian \eqref{eq:timeDependantHamiltonian} contains the term oscillating at the frequency $\Delta E/ \hbar$, to simplify the description of STIRAP evolution, it is necessary to consider a regime where the term effectively averages to zero.  To ensure that, the oscillation period, $T_{osc}=\hbar/{\Delta E}$, needs to be much shorter that the slowly-varying envelope characteristic times $T_{S,p}$
\begin{equation}
    T_{S,p} \gg \dfrac{\hbar}{\Delta E},
\end{equation}
which is manifested in Eq. \eqref{eq:separation} in the main text.
Under such conditions, the RWA Hamiltonian \eqref{eq:timeDependantHamiltonian} takes the form
\begin{widetext}
\begin{equation}
\small
    H = \hbar \left(\begin{array}{cccc}
     \delta & 0 & C_{1m_F ,1m_{F'}} \Omega_{p}(t) & C_{1m_F ,2m_{F'}} \Omega_{p}(t)\\
     0 & \delta & C_{2m_F ,1m_{F'}} \Omega_{S}(t) & C_{2m_F ,2m_{F'}} \Omega_{S}(t)\\
     C_{1m_F,1m_{F'}}^{*} \Omega_{p}(t) & C_{2m_F,1m_{F'}}^{*}\Omega_{S}(t) & 0 & 0 \\
     C_{1m_F,2m_{F'}}^{*} \Omega_{p}(t) & C_{2m_F,2m_{F'}}^{*}\Omega_{S}(t) & 0 & \Delta 
    \end{array}\right).
    \label{eq:hamiltonianSI}
\end{equation}
\end{widetext}
where the Stokes (pump) beam interacts only with the final (initial) state.

\subsubsection{Rubidium-87} 

In the main paper, we consider $^{87}$Rb atoms coupled with $\sigma^{+}$-polarized light, resonant at the D$_1$ line. For such a system, one can distinguish only two classes of Hamiltonians, (I) and (II).In case I, the Hamiltonian is given by
\begin{equation}
    H_{(\text{I})} = \hbar\left(\begin{array}{cccc}
     \delta & 0 & - \dfrac{\Omega_{p}(t)}{2\sqrt{6}} & - \dfrac{\Omega_{p}(t)}{2\sqrt{6}}\\
     0 & \delta &  \dfrac{\Omega_{S}(t)}{2\sqrt{2}} & \dfrac{\Omega_{S}(t)}{2\sqrt{2}}\\
     - \dfrac{\Omega_{p}(t)}{2\sqrt{6}} & \dfrac{\Omega_{S}(t)}{2\sqrt{2}} & 0 & 0 \\
     - \dfrac{\Omega_{p}(t)}{2\sqrt{6}} & \dfrac{\Omega_{S}(t)}{2\sqrt{2}} & 0 & \Delta 
    \end{array}\right).
    \label{eq:Hamiltonian1}
\end{equation}
One of the eigenstates of the Hamiltonian $H_{(\text{I})}$ is a dark state of the form $\ket{\mathrm{\delta}(t)} \propto \sqrt{3} \Omega_{S}(t) \ket{\tilde{\mathrm{i}}}+ \Omega_{p}(t) \ket{\tilde{\mathrm{f}}}$, with the  eigenenergy equal to $\hbar \delta$. The eigenenergies of other eigenstates are given by the roots of the Hamiltonian characteristic equation. It can be shown that if $\delta_{(\text{I})} = \Delta / 2$, there exist time $t_{0}$ at which $\Omega_{p}(t_{0}) = \Omega_{s}(t_{0})$, and one of the other eigenenergies is also equal to $\hbar \Delta / 2$, meaning that the dark state crosses with another eigenstate. The same applies to any case for which $C_{ie_{1}}=\pm C_{ie_{2}}$ and $C_{fe_{1}}=\pm C_{fe_{2}}$.

The case II Hamiltonian is given by
\begin{equation}
\small
    H_{(\text{II})} = \dfrac{\hbar}{2\sqrt{6}} \left(\begin{array}{cccc}
     \delta & 0 & - \Omega_{p}(t) & -\sqrt{3} \Omega_{p}(t)\\
     0 & \delta & \Omega_{S}(t) & \sqrt{3} \Omega_{S}(t)\\
     - \Omega_{p}(t) & \Omega_{S}(t) & 0 & 0 \\
     -\sqrt{3} \Omega_{p}(t) & \sqrt{3}\Omega_{S}(t) & 0 & \Delta 
    \end{array}\right).
\end{equation}
\normalsize
In this case, the dark state is given by $\ket{\mathrm{\delta}(t)} \propto  \Omega_{S}(t) \ket{\tilde{\mathrm{i}}}+ \Omega_{p}(t) \ket{\tilde{\mathrm{f}}}$, and the level crossing occurs at $\delta_{(\text{II})} = \Delta /4$. For the general case of class (II), i.e., $C_{ie_{1}}=\pm C_{fe_{1}}$ and $C_{ie_{2}}=\pm C_{fe_{2}}$, the crossing appears at $\delta_{(\text{II})} = \Delta\vert C_{ie_1}\vert^2/( \vert C_{ie_{1}}\vert^{2}+\vert C_{ie_{2}}\vert ^{2})$.
\vspace{15pt}

\subsubsection{Local adiabatic condition}


The Local Adiabatic Condition (LAC) follows the adiabatic theorem \cite{messiah1962quantum}, for which the necessary condition is that the eigenstates remain distinct from each other. The probability of a non-adiabatic transition from the dark state $\ket{\delta(t)}$ to an orthogonal eigenstate $\ket{\mathrm{y}(t)}$, is upper-bounded by
\begin{equation}
  P_{\ket{\delta} \rightarrow \ket{\mathrm{y}}} \lesssim \max_{t}\left\vert \dfrac{ \bra{\mathrm{y}(t)}\dfrac{d}{dt}\Big(\ket{\delta(t)}\Big)}{E_{\mathrm{y}(t)}/\hbar - \delta} \right\vert^{2},\label{eq:LACaa}
\end{equation}
where $E_{\mathrm{y}(t)}$ is the energy of the state $\ket{\mathrm{y}(t)}$ and $\delta$ is the energy of the dark state.
Importantly, if the above-mentioned energy separation is small, the evolution of the system must be slow enough so that the rate of change of the eigenstate in time is also small [see the derivative in the nominator in Eq.~\eqref{eq:LACaa}].

We are interested in the probability of transition of the system from the dark state $\ket{\delta}$ to an entire subspace $Q_{\delta}$ orthogonal to the dark state. In general, finding the formula for all eigenstates in a four-level system is complicated. Moreover, the inability to find a compact form of the state $\ket{\mathrm{y}}$ limits the usability of the formula~\eqref{eq:LACaa}. Nevertheless, the probability that the system transits from the dark to any other state can be upper bounded by the overlap of $\dfrac{d}{dt}\Big(\ket{\delta(t)}\Big)$ with the entire $Q_{\delta}$ subspace and dividing the result by the distance of the dark state to the closest eigenstate. This gives rise to
\begin{equation}
  P_{\ket{\delta} \rightarrow Q_{\delta}} \lesssim
  \max_{t} \left(
  \dfrac{ \left\vert
  \mathcal{P}_{Q_{\delta}} \left(
  \dfrac{d}{dt}\ket{\delta(t)} \right) \right\vert^{2}}{\min_{\mathrm{y}}\left\vert E_{\mathrm{y}(t)}/\hbar - \delta\right\vert^{2}} \right),
  \label{eq:16}
\end{equation}
where $\mathcal{P}_{Q_{\delta}}$ is the projection onto $Q_{\delta}$. It is convenient to extend the subspace $Q_{\delta}$ by a set of vectors $\{\ket{\tilde{\mathrm{e}}_1}, \ket{\tilde{\mathrm{e}}_2},\ket{\bar{\delta}}\}$.
\begin{equation}
  \ket{\bar{\delta}(t)} = \sin \vartheta \left(t\right) \ket{\tilde{\mathrm{i}}} + \cos \vartheta \left(t\right) \ket{\tilde{\mathrm{f}}}.
\end{equation}
Note that $\ket{\bar{\delta}(t)}$ is orthogonal to $\ket{\delta(t)}$ [given by Eq. \eqref{eq:DarkStateGeneric}]. The derivative $\dfrac{d}{dt}\Big(\ket{\delta(t)}\Big)$ is proportional to $\ket{\bar{\delta}(t)}$. Since the dark state $\ket{\delta}$ has no overlap with the excited states $\ket{\tilde{\mathrm{e}}_1}$ and $\ket{\tilde{\mathrm{e}}_2}$, the only non-vanishing contribution to the projection in the nominator of Eq. \eqref{eq:16} comes from the state $\ket{\bar{\delta}(t)}$, thus 

\begin{equation}
    P_{\ket{\delta} \rightarrow Q_{\delta}} \lesssim \max_{t}\left\vert\dfrac{\vert a \vert \dfrac{\vert \Omega_{p}(t) \dot\Omega_{S} (t) - \Omega_{S}(t) \dot\Omega_{p} (t)\vert}{\Omega_{p}^{2} (t)+ \vert a \vert^{2} \Omega_{S}^{2} (t)}}{ \min_{\mathrm{y}} \left\vert E_{\mathrm{y}(t)}/\hbar - \delta \right\vert }\right\vert^{2}.
    \label{eq:step_to_LAC}
\end{equation}

One would like to keep the probability small, $P_{\ket{\delta} \rightarrow Q_{\delta}} \ll 1$, during the entire STIRAP process. $P_{\ket{\delta} \rightarrow Q_{\delta}}$ is of the same order of magnitude as the right-hand side of Eq.~\eqref{eq:step_to_LAC} \cite{messiah1962quantum}, then it has to be much smaller than 1. This translates into the condition
\begin{equation}
  \text{min}_{\mathrm{y}\neq \delta} \left\vert E_{\mathrm{y}}/\hbar - \delta \right\vert \gg \vert a \vert \dfrac{\vert \Omega_{p}(t) \dot\Omega_{S} (t) - \Omega_{S}(t) \dot\Omega_{p} (t)\vert}{\Omega_{p}^{2} (t)+ \vert a \vert^{2} \Omega_{S}^{2} (t)},
\end{equation}
which was used in the main text [Eq.~\eqref{eq:localCondition}].

\bibliographystyle{apsrev4-1no-url}
\bibliography{STIRAPbiblio}
\end{document}